\documentclass[pdflatex,sn-mathphys-num]{sn-jnl}

\usepackage{graphicx}%
\usepackage{multirow}%
\usepackage{amsmath,amssymb,amsfonts}%
\usepackage{amsthm}%
\usepackage[title]{appendix}%
\usepackage{xcolor}%
\usepackage{textcomp}%
\usepackage{booktabs}%

\newcommand{\response}[1]{\textcolor{black}{#1}}

\usepackage{lineno}

\begin{document}

\title[Predictive Modeling for High Impact Active Learning Classrooms]{Predictive Modeling for High Impact Active Learning Classrooms}

\author*[1]{\fnm{Olive} \sur{Ross}}\email{ogr8@cornell.edu}

\author[2]{\fnm{Meagan} \sur{Sundstrom}}\email{sundstrommeagan@gmail.com}

\author[3]{\fnm{N.G.} \sur{Holmes}}\email{ngholmes@cornell.edu}

\affil*[1]{\orgdiv{Department of Astronomy}, \orgname{Cornell University}, \orgaddress{\street{410 Thurston Ave}, \city{Ithaca}, \postcode{14853}, \state{New York}, \country{United States}}}

\affil[2]{\orgdiv{Department of Physics}, \orgname{Drexel University}, \orgaddress{\street{3141 Chestnut St}, \city{Philadelphia}, \postcode{19104}, \state{Pennsylvania}, \country{Unites States}}}

\affil[3]{\orgdiv{Department of Physics}, \orgname{Cornell University}, \orgaddress{\street{410 Thurston Ave}, \city{Ithaca}, \postcode{14853}, \state{New York}, \country{United States}}}

\abstract{Over the past several decades, a large body of research has shown that undergraduate science students learn more and more equitably in active learning classrooms; however, the term “active learning” lacks definition and little research has examined which types and combinations of \response{associated} strategies are most effective. This study \response{uses} data from 69 courses, \response{3 scientific disciplines}, and 24 institutions to create a predictive model that maps classroom time spent on different activities to student conceptual learning. We find that \response{a model using four classroom variables -- time spent on lecture, group worksheets, \response{group} clicker questions, and student questions -- can make predictions that account for 49\% of the variance in effect sizes, as measured by $R^2$}. We identify one type of class that is  \response{predicted to have} exceptional student learning outcomes (effect sizes greater than 2): those that spend 10-20\% of class time on group worksheets, 20-40\% of class time on group clicker questions, and 10\% or more of class time on student questions. We also find that classes \response{that} do not utilize group worksheets have learning outcomes comparable to fully lecture classes. These results provide testable recommendations for future controlled studies to investigate effective active learning implementation in undergraduate physics courses.
}

\maketitle

\section{Introduction}
\label{intro}

Despite lecture being the most common form of instruction for undergraduate science courses [e.g., Refs. \cite{dancy_physics_2024, stains_anatomy_2018}], an extensive body of work has shown that classes in which instructors lecture for all of classroom time exhibit lower student learning \cite{freeman_active_2014}, higher student failure rates \cite{freeman_active_2014, prather_national_2009}, and less equitable student outcomes across demographic groups \cite{theobald_active_2020} than classes in which instructors utilize non-lecture methods. These non-lecture methods are collectively referred to as ``active learning" and encompass a wide variety of classroom activities, such as clicker questions, group worksheets, and think-pair-share activities \cite{stains_anatomy_2018, smith_classroom_2013}. Accordingly, governmental and institutional bodies recommend that science instructors incorporate active learning into their classrooms [e.g., Ref. \cite{olson_engage_2012}]. 

There are, however, myriad definitions of the term active learning, creating ambiguity around how to enact these recommendations \cite{lombardi_curious_2021}. There is also relatively little research pointing to which particular active learning strategies are most effective for student learning and for what portion of class time each strategy should be used to achieve these outcomes. Research studies that directly compare the effects of different active learning approaches on student learning outcomes are necessary to help instructors decide the types of active learning methods to implement and the best ways to implement them \cite{dancy_physics_2024}. 

Several existing studies have started to address this goal. One study \cite{sundstrom_relative_2026} compared physics student learning outcomes in four research-based pedagogical approaches, finding that the SCALE-UP (Student-Centered Active Learning Environment with Upside-down Pedagogies) pedagogy \cite{beichner_student-centered_2007} coincided with higher student learning gains than other commonly used active learning pedagogies. Another study \cite{weir_small_2019} related the extent to which different active learning activities were used during biology classes to student learning outcomes and found that spending more time on group worksheets was positively correlated with student learning. Other research has applied the ICAP (Interactive, Constructive, Active, and Passive) framework \cite{chi_active-constructive-interactive_2009} to distinguish between different types of engagement, finding that activities promoting higher levels of cognitive engagement are associated with better learning outcomes \cite{chi_icap_2014}. While this growing body of work has importantly started to disentangle the relative impacts of different active learning strategies, the impacts of different \textit{combinations} of these strategies has yet to be rigorously analyzed \response{across disciplines} in a way that produces testable predictions about student learning outcomes for future classes.

\response{Here we measure the impacts of these combinations by analyzing two types of} data: classroom activities and concept inventory scores. We characterize classroom activities using the Classroom Observation Protocol for Undergraduate STEM (COPUS) \cite{smith_classroom_2013}. The COPUS is a structured observation protocol that consists of 25 different codes, or student and instructor behaviors, and observers mark which codes occur in two-minute segments of class time. \response{In this work, we use the percentage of total coded intervals as a proxy for the percentage of class time spent on a given behavior.} \response{We confine our analysis to the eight COPUS codes identified by \citet{stains_anatomy_2018} as being the most descriptive of undergraduate STEM courses (see Section \ref{sec:COPUS} of the Methods for a more detailed discussion of code selection).} The COPUS has been used both as an internal assessment of teaching practices [e.g., \cite{bazett_course_2021}] and as a research tool [e.g., Refs. \cite{commeford_characterizing_2022,stains_anatomy_2018}]. \response{We use COPUS because it is the most common research tool for characterizing active learning instruction, with existing data publicly available.} 

Concept inventories are commonly used by discipline-based education researchers to measure student learning. Concept inventories are, usually, research-validated, multiple choice assessments that measure students' conceptual understanding of core concepts in a given topic (e.g., mechanics or introductory genetics). These assessments are typically administered both at the beginning of a course (pre-test) to determine students' baseline level of knowledge and at the end of a course (post-test) to measure how much students have learned from the received instruction. As in prior work [e.g., Ref. \cite{sundstrom_relative_2026}], \response{we use effect sizes between matched students' pre-test and post-test scores within a single class to quantify student learning. Notably, this is distinct from using effect sizes to compare outcomes between classes, as in, e.g., Ref. \citep{freeman_active_2014}.} We particularly use \response{Hedges' \textit{g} \citep{hedges_statistical_1982}, a small sample adjusted variation of Cohen's $d$,} as this measure can be compared across different concept inventories, class sizes, and student populations \cite{burkholder_examination_2020, nissen_comparison_2018, sundstrom_relative_2026} \response{(see Section \ref{sec:CIs} of the Methods for a more detailed discussion of this metric).}

We leverage both publicly available data of COPUS observations and student concept inventory scores from three extant studies \cite{sundstrom_relative_2026, connell_increasing_2016, weir_small_2019} and new data collected by the authors---spanning 24 American and Canadian institutions, 69 undergraduate \response{physics, astronomy, and biology} courses, and 40 unique instructors in total---in this analysis. This dataset involves more than 11,000 students in classes with sizes ranging from 11 to 576 students. \response{A detailed breakdown of the institutions and classes represented in this dataset can be found in Table \ref{tab:data_sources} in the Methods}. We use these data to develop quantifiable and testable predictions for which combinations of active learning strategies optimize student learning gains in undergraduate physics and biology courses. Notably, we move beyond descriptive and/or explanatory analyses conducted in prior studies of the effect of active learning strategies and instead utilize a predictive framework as is common in many other science fields \cite{shmueli_explain_2010, aiken_framework_2021}. \response{Rather than approximating the extent to which an instructional strategy correlates with student learning gains, as in explanatory modeling, a predictive model creates testable estimates for the student learning gains that would result from class time spent on different active learning strategies.}

\section{Results}
\label{sec:results}
We use \response{weighted least squares (WLS)} regression to train a model that maps the fraction of two-minute class intervals in which specific COPUS codes were present (\response{a proxy for} the fraction of class time spent on a given activity) to the concept inventory effect size in that class. \response{We iteratively perform feature selection, retaining the 10 combinations of features with the lowest AIC values. We then} apply bootstrap sampling to create a suite of functions rather than a single function, such that the model outputs a distribution of predicted effect sizes rather than a single predicted effect size. The final model prediction is then taken as the \response{AIC-weighted} mean of this distribution and the prediction uncertainty (due to sample bias) is taken as the standard deviation of the distribution. 

\begin{figure}[b]

    \centering
    \includegraphics[width=0.7\linewidth]{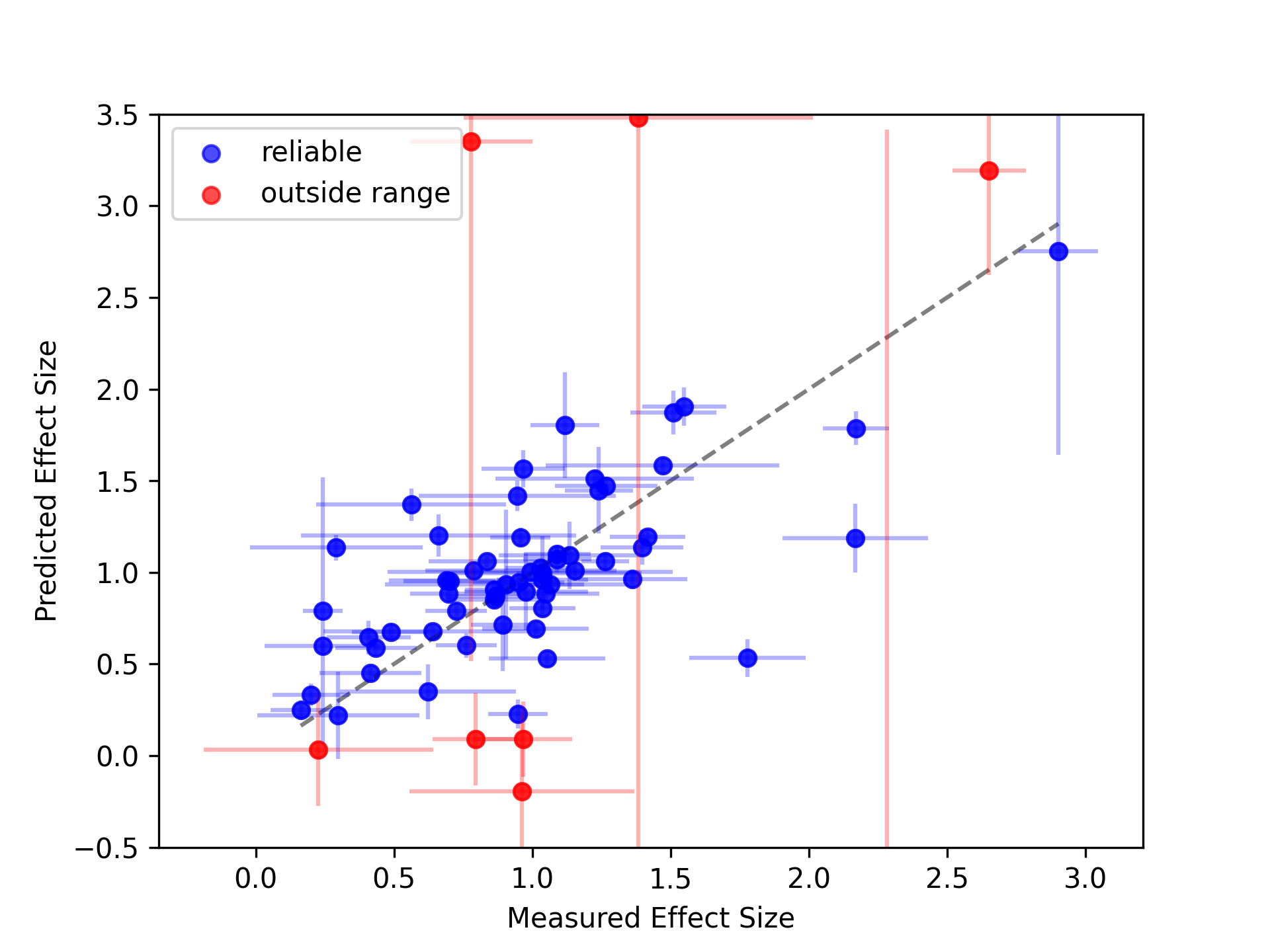}
    
    \caption{Results of Leave-One-Out tests for all 69 classes.  The dashed line represents $y=x$. Each class is removed from the dataset, the model is trained, and a prediction is made for the left-out class. The vertical error bars represent model uncertainty on the predictions, while horizontal error bars represent the measurement uncertainty of the effect sizes, as defined in Section \ref{sec:CIs} of the Methods. Model predictions outside the range of measured effect sizes in the data set are shown in red \response{and excluded from goodness of fit metrics}.}
    \label{fig:LOO}

\end{figure}

\response{We assess models incorporating various combinations of COPUS codes and varying model complexities, and ultimately select a model that includes up to fourth order terms of  $Lec$ (instructor lecturing), $CG$ (students working on clicker questions in groups), $WG$ (students working on worksheets in groups), and $SQ$ (a student asking a question in front of the whole class; see Sections \ref{sec:COPUS} and \ref{subsec:Modeling} in the Methods for further model selection details).} We perform leave-one-out tests for each class in the dataset to test for overfitting. Leave-one-out tests involve removing one class from the dataset, \response{training the model (feature selection, bootstrapping, etc.)}, and then using that model to make a prediction on the left-out class. This allows us to test how accurately the model is able to make predictions for classes outside of its training data.
The selected model is limited by the data it was trained on; therefore, \response{we define any predicted effect size outside the range of those effect sizes present in the training data as unreliable.} \response{We remove the 9 predictions outside of the measured range before computing prediction metrics because we do not include these predictions in our analysis.} The mean absolute error between predicted and measured effect sizes in the leave-one-out tests for predictions falling within the measured range is \response{0.26}, meaning that the average difference between measured and predicted effect sizes is $\approx$ 10\% of the full range of effect sizes represented in the data set. We compare measured and predicted effects sizes along with prediction error in Figure \ref{fig:LOO}. \response{The $R^2$ value of the model is 0.49, meaning that roughly 49\% of the variance in effect sizes in this dataset is accounted for by the model.} \response{As discussed in Section \ref{subsec:bootstrap} and shown in Figure \ref{fig:LOO_N_Field} of the Methods, we find no significant variations in predictions accuracy across different scientific disciplines or class sizes.}

\begin{figure}[htbp]

    \centering

    \includegraphics[width=1\linewidth]{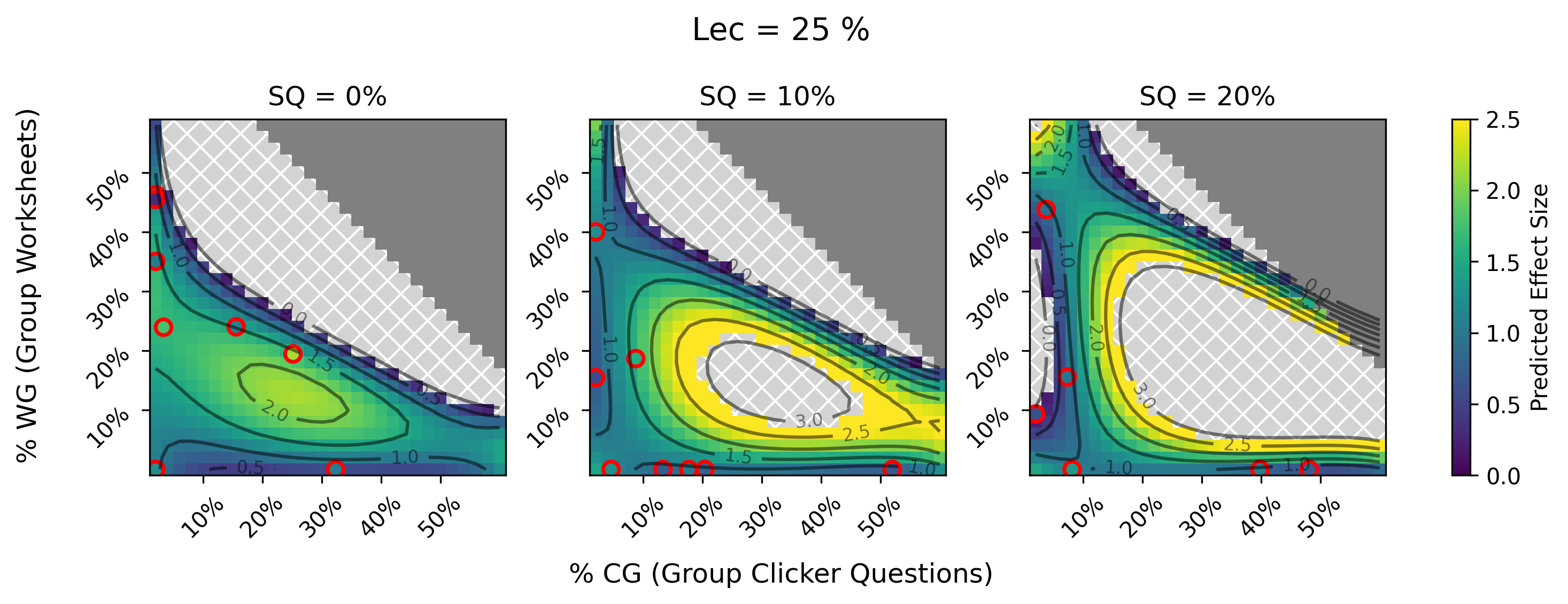}
    \includegraphics[width=1\linewidth]{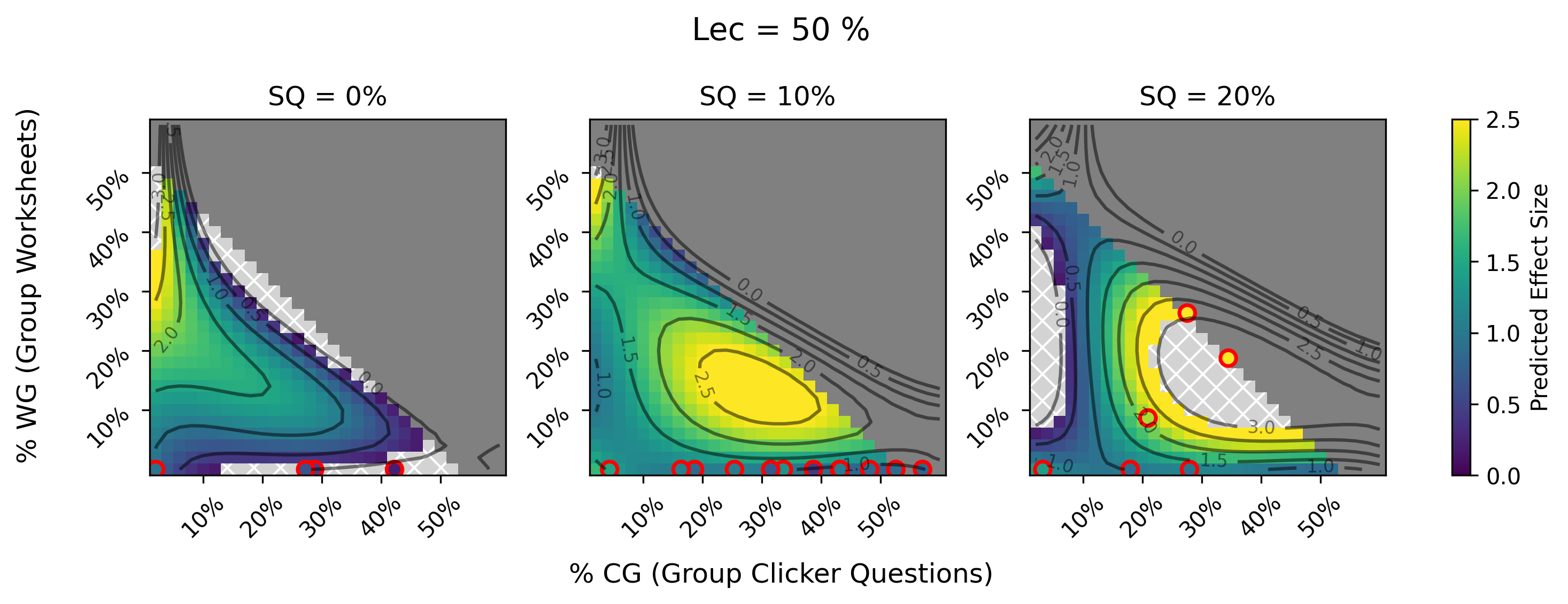}
    \includegraphics[width=1\linewidth]{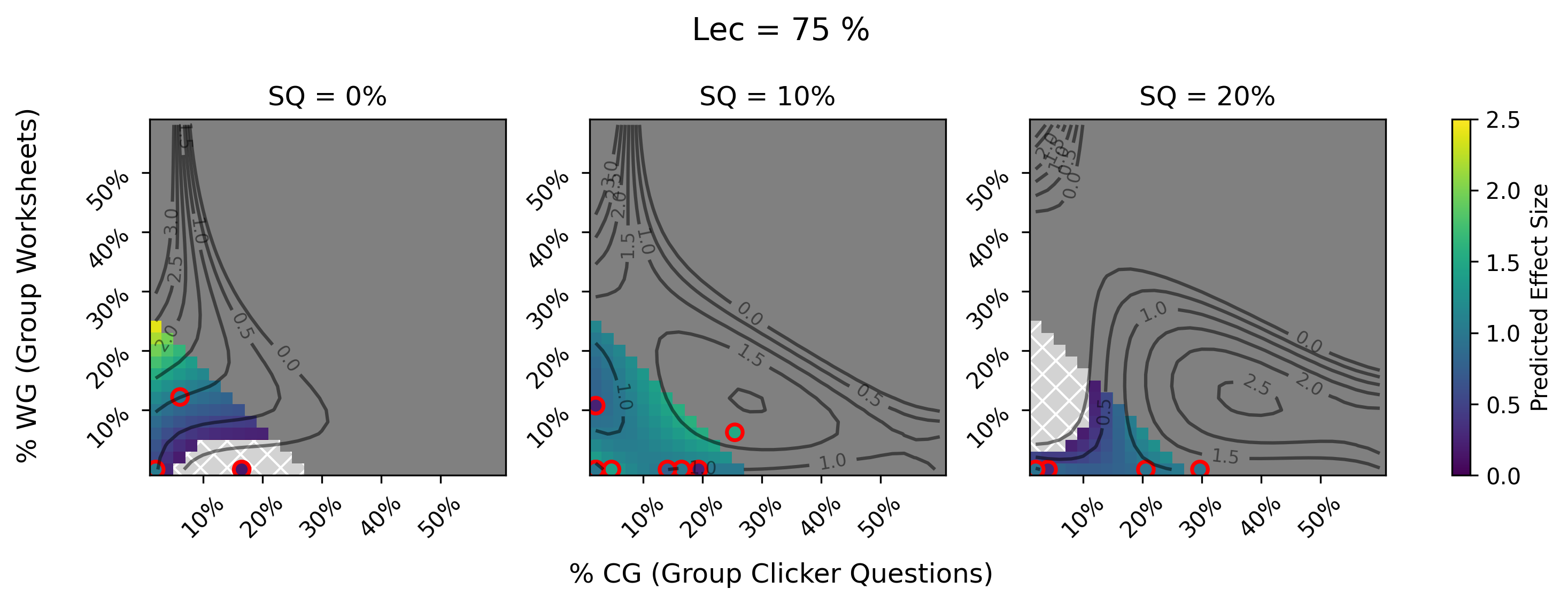}
    \caption{Heatmaps of predicted effect sizes from various combinations of COPUS code data by fraction of 2-minute intervals of class time coded as group worksheets (WG), group clicker questions (CG), student questions (SQ), and lecture (Lec). The heatmaps are overlaid with the measured data, represented by red-lined circles with fill colors corresponding to measured effect size. \response{Measured data is binned to the nearest value of SQ and Lec. The first, second, and third rows represent 25\%, 50\%, and 75\% of 2-minute intervals coded as lecture respectively. The first, second, and third columns represent 0\%, 10\%, and 20\% respectively of 2-minute intervals coded with SQ. The x-axis represents CG and the y-axis represents WG. The hatched shaded regions represent predictions outside the range of measured effect sizes, while the dark gray shaded regions represent regions of parameter space that are unlikely to be realizable ($WG + CG + Lec > 1$).}}
    \label{fig:heatmaps}
    
\end{figure}

\response{We trained the selected model on all 69 classes} and used the model to predict the effect sizes for simulated COPUS data that span the parameter space of the existing data. \response{These predictions are shown alongside real class data in Fig. \ref{fig:heatmaps}. } \response{The modeled effect sizes show several key predictions about the impacts of combinations of instructional strategies on student learning:}
\begin{enumerate}
    \item Classes that use no group worksheets are predicted to have consistently small effect sizes, even when other active learning strategies (e.g., group clicker questions) are used \response{(y = 0 on all panels in Fig. \ref{fig:heatmaps}).}
    \item \response{Classes that use no group clicker questions have a wide range of predicted effect sizes, depending on the prevalence of other instructional activities (e.g., panel 2 row 2 of Fig. \ref{fig:heatmaps}).}
    \item Classes that spend 10-20\% of class time on group worksheets, 20-40\% of class time on group clicker questions, and more than 10\% of class time on student questions ($\approx$2 student questions per hour) \response{are predicted to have} large effect sizes (third, fourth, and fifth panels of Fig. \ref{fig:heatmaps}).

\end{enumerate}

\section{Discussion}

Our analysis identifies specific types and combinations of active learning strategies that \response{are predicted to} coincide with large student learning gains in undergraduate science courses. In this section, we discuss our findings in light of previous work and, where relevant, suggest avenues for future work. First, we find that in-class group worksheets are strong predictors of student learning. This result agrees with the findings of Weir and colleagues \cite{weir_small_2019} \response{despite the different analysis methods used in that work and the expansion to multiple disciplines in our work.} As group worksheets are inherently interactive activities, this result also aligns with the ICAP \cite{chi_active-constructive-interactive_2009} framework's assertion that constructive and interactive activities lead to the highest learning gains. \response{Importantly, the predictive model suggests classes need only dedicate 10\% of class-time to group worksheets, in combination with other active learning strategies, to achieve large learning gains.}

Second, our model indicates that student questions are \response{also predictive of student learning. All classes spending more than 10\% of class time on both group clicker questions and group worksheets are predicted to have larger learning gains if students are asking more questions. } While past work has investigated the impact of student questions on the asker and the instructor \cite{chin_students_2008, rowe_wait_1986}, \response{here we uniquely observe} the impact of individual, in-class, student questions on the learning outcomes of the whole class. \response{We also find that more class time spent on student questions is correlated with a decrease in the width of the distribution of student scores from pre-test to post-test (see Fig. \ref{fig:copus_dists} in the Methods), potentially indicating that student questions can decrease the variation in student post-test scores. The mechanisms through which a single student's question impacts the whole class's learning should be explored in future work, for example through student and/or instructor interviews.}

Third, there is one combination of active learning strategies which is \response{predicted to result in the highest effect sizes} and should therefore be tested in controlled studies. Classes that spend 20-40\% of class time on group clicker questions, 10-20\% of class time on group worksheets, and 10\% or more of class time on student questions are consistently predicted to have high learning effect sizes, even when all other class time is spent lecturing. \response{These predictions align with the outcomes of the real classes in the dataset that fall within this range (sixth panel of Fig. \ref{fig:heatmaps}}). \response{We note that the measured data on which these predictions are made is reported as percentages of two-minute intervals in which a given code was observed, which may be a slight overestimate of the true class time spent on these activities.}

Lastly, the results indicate that less class time spent on lecture is not \response{inherently better for student learning, as suggested by some prior work [e.g., Ref. \cite{dancy_physics_2024}]. While it is true that classes that are almost entirely lecture-based consistently yield lower learning outcomes, according to our model, classes spending 50\% of class time on lecture can achieve comparably high effect sizes to classes spending 25\% of class time on lecture.} \response{Most distinct from prior work}, we find that classes which rely on group clicker questions as the only form of active learning (that is, no group worksheets) have consistently low effect sizes, comparable to the effect sizes seen in fully lecture classes. This combination of results is particularly surprising given the abundance of research studies evaluating the effectiveness of clicker-based pedagogies, such as Peer Instruction, as compared to traditional lecture courses [e.g., Refs. \cite{crouch_peer_2001, bojinova_teaching_2013}]. \response{Future work should seek to disentangle this apparent contradiction, such as by investigating the effects of fidelity of implementation or timescales of assessment on the measured gains in student learning.}

Several limitations of this study also prompt future investigations. An undergraduate science class has many more variables than are captured in our model. First, while there are many potential metrics and outcomes on which to assess the quality of an undergraduate science course (e.g., problem solving skills, effective collaboration, self efficacy, critical thinking), in this study we focus solely on conceptual learning. Future work should examine the differential impact of active learning strategies on other student outcomes. We also do not account for activities that occur outside of regularly scheduled lecture time (such as homework, quizzes, projects, laboratory sections, discussion sections, or time spent studying) or for in-class variables not captured by the four COPUS codes (CG, SQ, Lec, and WG) included in this study. \response{Our model explains just under 50\% of the variance in the data, which motivates investigation into other instructional activities. }Even within the four variables considered in this work, the sparsity of our data in certain regions of this parameter space \response{also} limits our ability to make reliable predictions \response{for certain combinations of the strategies. Thus, future studies that examine combinations of instructional strategies representing the sparsely populated areas would provide further tests of the predictive power of the model. A larger dataset would also allow for more complex models to be considered by lowering the risk of overfitting.} 

Our COPUS data also represent average rates of code occurrence over two to nine observed class sessions per course. While a sample of two to three classes is generally considered a representative sample of classroom activities \cite{lund_best_2015,stains_anatomy_2018}, the particular class sessions observed in our dataset may not represent typical or average classroom behaviors across the entire course. Additionally, because COPUS coding occurs in two-minute intervals, an occurrence of a code of any duration within a given \response{two-minute }interval is marked as being present for the full interval. This feature of the protocol may lead to an overestimation of class time spent on \response{specific activities}. Future work should consider using continuous coding instead of two-minute intervals to capture more accurate fractions of class time spent on different activities. \response{Additionally, it may aid future work to consider the sub-coding of COPUS codes. For example, the $PQ$ code is marked both when an instructor verbally asks a question that one student answers \textit{and} when students are actively working on worksheets or other assigned group work. Distinguishing between different activities that fall under the same code may improve the variation captured during modeling.}

Notably, this work focuses on the statistical accuracy of model predictions and not the theoretical underpinnings or causality of the relationships represented by the model. Future work should evaluate whether the presence of these activities directly impacts student learning, or if they serve as proxies for other instructional or interpersonal variables. \response{For example, while we have both measured in our dataset and predicted with our model that group worksheets tend to have a larger impact on effect sizes than group clicker questions, we have not investigated the underlying mechanisms of this observation. One possibility is that clicker questions are generally less involved than worksheet questions (e.g., multiple choice instead of open ended problems), and the driver of student learning is complexity not mode of delivery. Future studies may test this and other possibilities by conducting think-aloud interviews with students doing each type of activity about the same topic to better understand how students develop understanding when completing different types of active learning tasks.}

\response{Lastly, we note that there may be some selection bias} effects both for instructors who volunteered their classes to participate in these studies (e.g., instructors who are discipline-based education researchers themselves may be more likely to volunteer and may exhibit larger student learning gains than other instructors \cite{andrews_active_2011}) and for students who completed both the pre-test and post-test concept inventories (e.g., high-performing students are more likely to participate in research studies than low-performing students \cite{nissen_missing_2019}).
Future work should examine these and other factors that may impact the model results, such as by further comparing the model's prediction accuracy across different populations of students, disciplines, course levels, and other contextual and demographic variables.

\section{Methods}
\label{sec:methods}
\subsection{Data}
\label{subsec:Data}
There are two sources of data in this study: those sourced from extant studies and those collected for the current study. We use extant data from three different published works \cite{sundstrom_relative_2026, connell_increasing_2016, weir_small_2019}. We identified these works by first searching for all papers that have cited Ref. \cite{smith_classroom_2013}, the paper that introduces the COPUS, and then identifying those that collected both COPUS and concept inventory data. From these works, we selected those with publicly available data containing the \response{fraction of total coded intervals} that each COPUS code was marked as present and concept inventory means and standard deviations. We also contacted the authors of papers without publicly available data; however, none of those authors were able to provide the required data for our analysis. \response{The nature of the dataset limits our observations to three disciplines: physics, astronomy, and biology.} To supplement these data, we also collected COPUS and concept inventory data from 12 introductory physics and astronomy class sections at a private, R1 university in the United States. The full dataset represents 69 class sections, 24 American and Canadian Universities, more than 11,000 enrolled students, and 6,680 matched student concept inventory pre-test and post-test scores. Descriptive statistics of all included classes are shown in Table \ref{tab:data_sources}.

\begin{table}[htbp]

\centering
\caption{The research studies from which we draw data and the course and institution types represented by these data. Values in the rightmost column represent the total number of each type represented in the final predictions.}
\begin{tabular}{llccccl}

\hline\hline
& &  \cite{connell_increasing_2016}&\cite{weir_small_2019}& \cite{sundstrom_relative_2026} & Current&Total\\
& &  & & \ & Study&\\\hline
 Total classes& & 2& 29& 26& 12&69\\ \hline

Highest degree granted \\ at institution& Associate's&  0&0& 1& 0&1\\
 & Bachelor's&  0&0& 1& 0&1\\
 & Master's&  0&0& 9& 0&9\\
 & PhD&  2&29& 15& 12&58\\\hline 

Research designations\\ of institution
& R1&  0&29& 8& 12&49\\
 & R2& 0& 0& 5& 0&5\\
 & RCU& 2& 0& 4& 0&6\\
 & None& 0& 0& 9& 0&9\\\hline

Public/private \\institutions & Public&  2&29& 16& 0&47\\
 & Private& 0& 0& 10& 12&22\\ \hline
Disciplines                           & Physics&  0&0& 24& 10&34\\
 & Astronomy& 0& 0& 2& 2&4\\
 & Biology& 2& 29& 0& 0&31\\ \hline
Class sizes                           & Range & 172 - 182  & 12 - 551 & 11 - 576 & 27 - 260 &11 - 576 \\
& Median &  177 & 225 & 45 &  146 &168\\ 
& Total &  354 &  6556& 2637& 1949  & 11,172\\ \hline
Response rates                          & Range &  83-87\% & 21-100\% & 40-98\% & 41-89\%& 21-100\% \\
& Median & 85\%  & 61\% & 75\%&  68\%&66\% \\\hline
Class levels                         & Lower division&  2&22& 26& 12&62\\ 
 & Upper division& 0& 7& 0& 0&7\\ \hline\hline
\end{tabular}

\label{tab:data_sources}
\end{table}

\subsection{Quantifying Student Learning with Concept Inventories}
\label{sec:CIs}
We measure student learning using the effect size \response{Hedges' \textit{g} \citep{hedges_statistical_1982}}, a standardized mean difference between matched pre-test scores and post-test scores on concept inventories \response{within a single class}, defined as:

\begin{equation}
    g = \dfrac{\mu_{post} - \mu_{pre}}{\sqrt{\frac{\sigma_{pre}^2 + \sigma_{post}^2}{2}}}c(m),
\end{equation}

where $\mu_{pre}$ and $\mu_{post}$ are the mean student scores on the pre-tests and post-tests, respectively, $\sigma_{pre}$ and $\sigma_{post}$ are the standard deviations of the pre-test and post-test scores, respectively,  and $c(m)$ is the small sample correction approximated as
\begin{equation}
    c(m) = 1 - \dfrac{3}{4m - 1},
\end{equation}
where $m$ is defined as

\begin{equation}
    m = 2N - 2
\end{equation}

and N is the number of matched student scores used to calculate the effect size. \response{As in Ref. \citep{hedges_statistical_1982}, we calculate the variance of the calculated effect size as}

\begin{equation}
    \sigma_g^2 =  \dfrac{c(m)^2m(1 + \frac{N}{2}g^2)}{N(N-2)} - g^2
\end{equation}

\begin{figure}[htbp]

    \centering
    \includegraphics[width=0.5\linewidth]{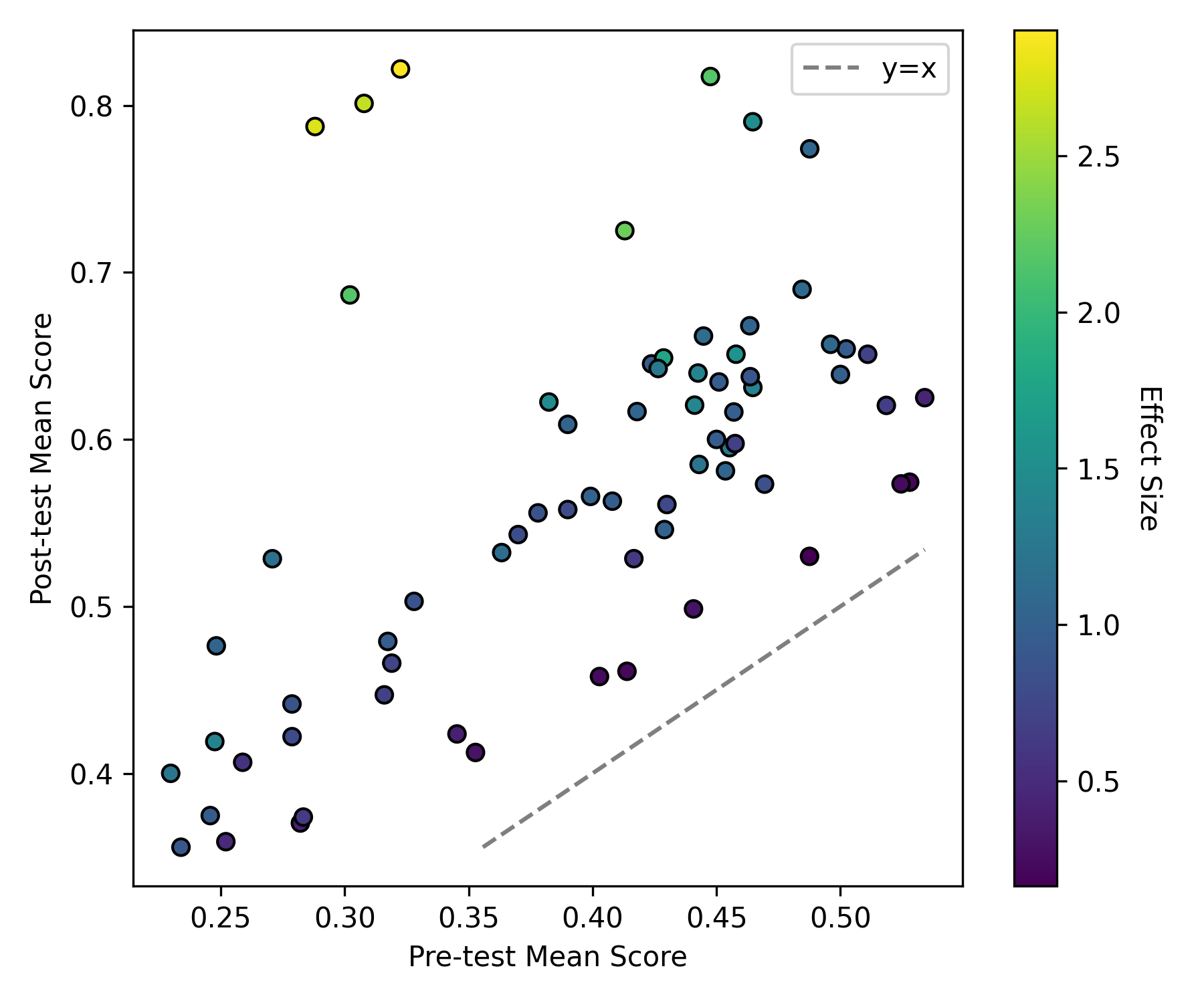}
    
    \caption{Mean post-test scores as a function of mean pre-test scores colored by effect size. The dashed line marks $y=x$, so a class falling on that line would have zero learning gain. Pre-test standard deviations range from 0.07 to 0.28, while post-test standard deviations range from 0.12 to 0.31.}
    \label{fig:pre_post}
    
\end{figure}

\begin{figure}[htbp]

    \centering
    
    \includegraphics[width=.5\linewidth]{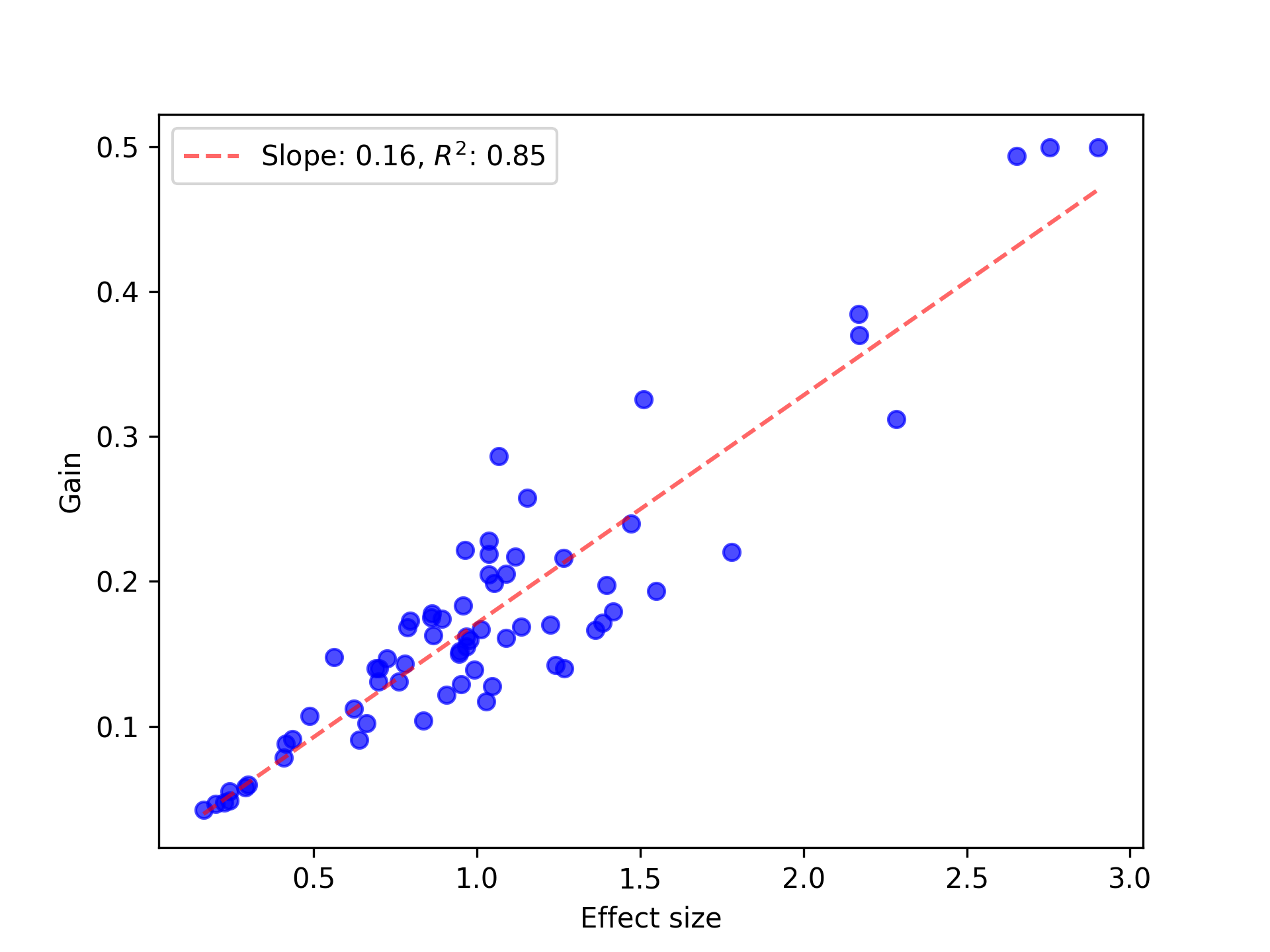}
    \caption{Relationships between effect sizes and gains (mean post-test scores minus mean pre-tests scores). The dashed red line denotes a best fit line with a slope of 0.16, indicating that an effect size of one is comparable to a gain of 16\%.}
    \label{fig:gains}
    
\end{figure}

\response{which allows us to perform variance-weighted least squares regression.} This metric provides a dimensionless measure of student learning in each course, allowing us to \response{compare student learning outcomes} across a range of disciplines, courses, student populations, and concept inventories. The ability to compare scores across concept inventories is particularly important in this work, with 28 different concept inventories represented across our dataset. Fig. \ref{fig:pre_post} shows the relationships between concept inventory mean pre-test scores, mean post-test scores, and effect sizes present in the data.  The nature of this work necessitates the use of effect size as a summary statistic, however, as demonstrated in Fig. \ref{fig:gains}, effect size is strongly correlated with gain. This may aid in interpretation for readers with more familiarity with gain as a summary statistic. 

We removed three classes from the dataset because they had pre-test scores over 60\% and high pre-test scores limit the range of possible learning gains, creating ceiling effects that lead to non-normal distributions of scores. \response{These classes are not included in the 69 classes detailed in Table \ref{tab:data_sources}.}

\subsection{Quantifying Classroom Activities with the COPUS}
\label{sec:COPUS}

The COPUS data were all collected either through live classroom observations or video recordings. To apply the COPUS to a live or video-recorded classroom observation, a trained coder indicates whether or not each COPUS code was present during each two-minute time interval \cite{smith_classroom_2013}. The COPUS code quantities reported here are the percentage of total two-minute time intervals observed in which the given code was present. Multiple COPUS codes can be applied in each time interval; therefore, the sum of all COPUS code quantities is more than one. 

\begin{figure}[htbp]

    \centering
    \includegraphics[width=1\linewidth]{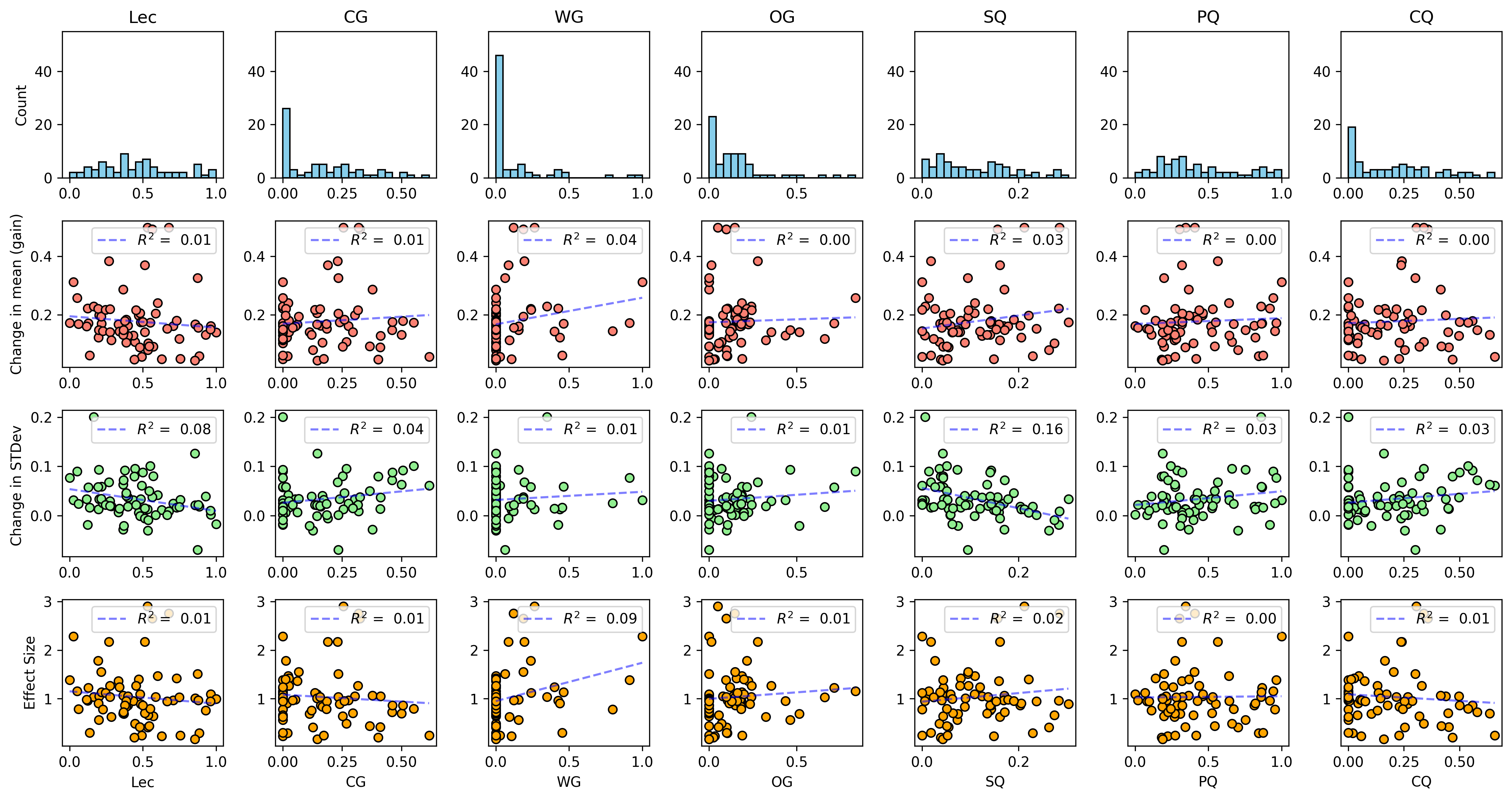}
    
    \caption{Distribution of classes with respect to COPUS values (row 1), plots of COPUS code values against the difference between mean post-test scores and mean pretest scores, or gain  (row 2), plots of COPUS code values against the difference between the standard deviation of posttest scores and the standard deviation of pretest scores (row 3), and plots of COPUS code values against effect sizes (row 4).}
    \label{fig:copus_dists}
    
\end{figure}

\response{\citet{stains_anatomy_2018} identifies 8 COPUS codes that ``were observed [in undergraduate STEM classes] with adequate heterogeneity, were not highly correlated with each other, and were likely to be key strategies in active or non-active learning environments". These codes include Lec (instructor lecturing), PQ (instructor posing a question to the class), CQ (instructor posing a clicker question), 1o1 (instructor having a one-on-one extended conversation with a student), WG (students working on worksheets in groups), CG (students working on clicker questions in groups), SQ (student asking a question in front of the whole class), and OG (students working on other assigned group activity). Our modeling, therefore, started with this contained list of COPUS codes. We removed the COPUS code 1o1 because 
all COPUS coding in the \citet{sundstrom_relative_2026} dataset and some coding in the newly collected dataset are done using classroom recordings. These recordings often do not capture the full classroom setting, making it infeasible to consistently determine the duration of an instructor's interactions with specific students during group work. This code is also very rarely applied in the existing data, as mentioned in Ref. \cite{sundstrom_relative_2026}, lacking the variation needed to make meaningful predictions from its presence.} 
Figure~\ref{fig:copus_dists} shows the prevalence of values for each of the remaining seven COPUS codes in the dataset as well as their relationships against the raw difference in concept inventory score means (gain), the change in concept inventory score standard deviations, and effect sizes.

\subsection{Modeling}
\label{subsec:Modeling}
In this section, we describe our development of a predictive model that maps COPUS codes to effect sizes. Source code is available at \url{https://doi.org/10.5281/zenodo.23819570}. \response{We contrast the goals of this predictive approach to traditional explanatory modeling in the main text. Here, we distinguish some analytical choices. For example, traditional explanatory modeling is typically confined to first and sometimes second order terms because the goal is to interpret the weights of the model. Higher order terms, particularly interaction terms, become hard to interpret and therefore explain. In predictive modeling approaches, the goal is to accurately make predictions on new data and not to explain the structure of the model. This approach removes the need to confine the model to first and second order terms, and allows future work to focus on testing the outputs of the prediction, not the explanatory mechanisms.} Indeed, the relationship between classroom activities and student learning outcomes is complex. Some activities may have simple linear relationships with learning (e.g., lecture time is negatively correlated  with effect size) while others may have more complicated relationships (e.g., effect size peaks at 20-30\% of class time spent on group clicker questions). \response{The possibility of these non-linear effects means we should consider models with varying complexity, such as those shown in Table \ref{tab:models}.} First order models consider only first order terms (e.g., $Lec$, $WG$), second order models consider first and second or terms as well as second order cross terms (e.g., $Lec * WG$, $Lec^2$), third order models add all third order terms and interaction (or cross) terms (e.g., $Lec * WG * SQ$, $Lec^2 * WG$, $Lec^3$), and fourth order models add the quartics of each variable (e.g., $Lec^4$, $WG^4$).

\begin{figure}
    \centering
    \includegraphics[width=0.7\linewidth]{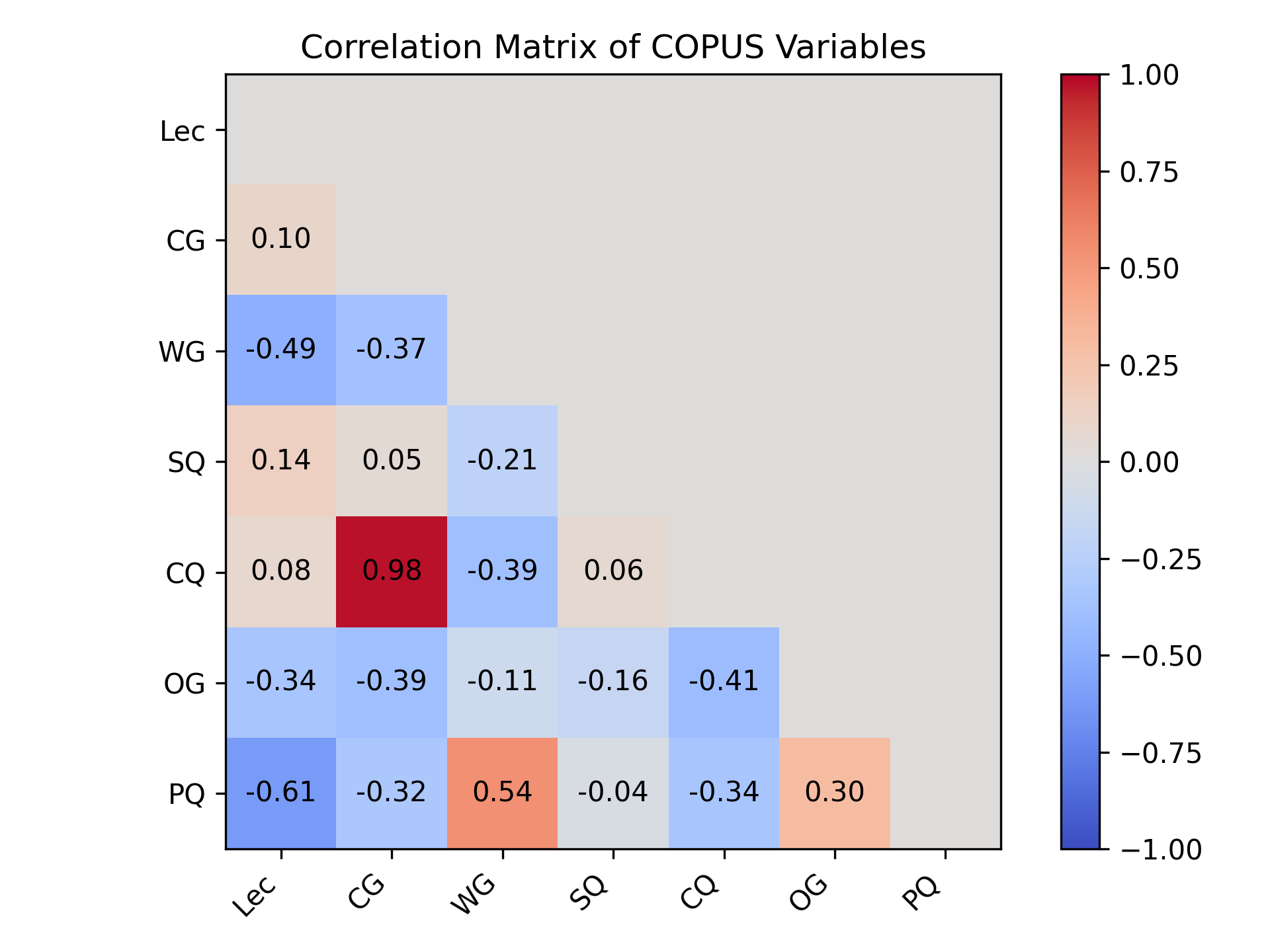}
    \caption{\response{Correlation matrix of the seven COPUS codes considered in this analysis.}}
    \label{fig:cov}
\end{figure}

\response{As shown in Fig.~\ref{fig:cov}, our analysis must consider the correlations between instructor and student codes. The codes CQ and CG are almost exactly correlated, indicating co-occurrence. The only times when these codes are not marked in the same interval are the few times students are explicitly asked to work on a clicker question independently, or when an instructor is posing the clicker question but has not yet asked students to work on it. For this reason, we only consider models that use one or the other of these codes. Similarly, PQ is positively correlated with both WG and OG. By definition, PQ is marked anytime the student codes OG or WG are marked \citep{smith_classroom_2013}, and so we consider models that use PQ or models that use WG and OG. Several codes are also negatively correlated, which is likely an inherent feature of the data; the total time spent on COPUS codes is a fixed quantity, and more time on one code necessarily means less time on another code. Because we do not include all possible classroom activities, the sum of codes included in the models is not a fixed quantity. Furthermore, our feature selection process (outlined below) further constrains the set of variables in our models, meaning the correlations between terms is less concerning in the technical modeling (though should still be considered in the subsequent interpretations and future studies). All combinations of codes considered can be found in Table \ref{tab:models}.}

\begin{table}

\begin{tabular}{|l|l|l|l|l|l|}
\hline
\textbf{Codes Included}              & \textbf{Highest }& \textbf{LOO} & \textbf{LOO} & \textbf{Number} & \textbf{Minimum} \\ 
\textbf{}              & \textbf{Order}& \textbf{$R^2$} & \textbf{MAE} & \textbf{Removed} & \textbf{AIC} \\ 
& \textbf{Terms} & & & &\\ \hline
\textbf{Lec, CG, WG, OG, SQ}         & 4th                          & .35& .32&  9& 10.79\\ \hline
\textbf{Lec, CQ, WG, OG, SQ}         & 4th                          & .12& .42&  4& 28.20\\\hline
 \textbf{Lec, CG, PQ, SQ}& 4th                          & .08& .47& 5&94.73\\\hline 
\textbf{Lec, CG, WG, SQ}& 4th                          & .49& .26&  9& 35.87\\ \hline
\textbf{CG, WG, OG, SQ}              & 4th                          & .17& .37&  5& 27.27\\ \hline
\textbf{Lec, CG, OG, SQ}             & 4th                          & .23& .41&  3& 87.61\\ \hline
\textbf{Lec, CG, WG, OG}             & 4th                          & .29& .36&  3& 48.62\\ \hline
\textbf{Lec, CG, {[}WG + OG{]}, SQ}  & 4th                          & .12& .44&  2& 72.56\\ \hline
\textbf{Lec, {[}CG + OG{]}, WG, SQ}  & 4th                          & .23& .39&  7& 62.16\\ \hline
\textbf{Lec, {[}CG + WG{]}, OG, SQ}  & 4th                          & .28& .33&  10& 20.79\\ \hline
\textbf{Lec, {[}CG + WG + OG{]}, SQ} & 4th                          & .07& .46& 3& 93.11\\ \hline
\textbf{Lec, CG, WG, OG, SQ}         & 3rd                          & .15& .38& 9& 29.05\\ \hline
\textbf{Lec, CG, WG, SQ}             & 3rd                          & .24& .34& 4& 33.14\\ \hline
\textbf{Lec, CG, WG, SQ}             & 2nd                          & .30& .34& 3& 62.34\\ \hline
\textbf{Lec, CG, WG, SQ}             & 1st                          & .09& .46& 0     & 121.47\\ \hline
\end{tabular}

\caption{Descriptive statistics of models with differing complexities and input features. LOO $R^2$ and LOO MAE describe the $R^2$ value and mean absolute error between measured effect sizes and effect sizes predicted in the leave-one-out tests. These values are computed after predicted effect sizes outside the range of measured effect sizes are removed, to simulate how accuracy would be assessed on new data. "Number removed" is the number of predictions that were removed before calculating the prior two values. Minimum AIC describes the minimum value of AIC in training, describing only how well the training data fits to the model. Therefore a low AIC may indicate overfitting.}
\label{tab:models}
\end{table}


For each model shown in Table \ref{tab:models}, we used the weighted least squares (WLS) module from the \textsc{statsmodels} \cite{seabold_statsmodels_2010} package in Python to fit weights on all features. We then computed the small sample corrected Akaike Information Criterion (AIC), \response{which is a metric that rewards goodness-of-fit by maximizing a likelihood function and penalizing model complexity. After computing the AIC for that fit, one feature was removed, the fitting algorithm was run again, and the AIC was computed. If the AIC decreased by less than 2 when that feature was removed, meaning the model fit improved or did not meaningfully decrease \citep{burnham_model_2004}, then that feature was removed from the model.} This process was repeated for all features to select features for the final model. The order in which this feature removal is performed has an impact on the features that are ultimately selected. For this reason, this entire feature selection process was repeated $10\times$ the number of model terms (as suggested by \citet{burnham_model_2004}) with different feature orders. The feature sets with the 10 lowest associated AIC values were retained and their associated predictions were weighted relative to the lowest AIC values, which are shown in Table \ref{tab:models}. 

\response{After feature selection, we performed bootstrap sampling.} The dataset used to train our model is limited (i.e., it does not include every possible type of class); therefore, the predictions we are able to make are limited. To quantify how sensitive the model predictions are to sample bias, we performed bootstrap leave-one-out testing. For each  set of feature weights within each leave-one-out test, we randomly selected 30 different subsets of 85\% of the data and trained the model on each of those subsets. The feature selection and bootstrapping process was repeated 10 times, meaning that each model prediction consists of a distribution of 300 predictions. The model predictions are weighted by their AIC values relative to the lowest AIC value to account for the likelihood of each model being the true underlying model as recommended in \citet{burnham_model_2004}. The final prediction is taken as the AIC-weighted mean of that distribution and the model uncertainty on that prediction is defined as the standard deviation of that distribution. 

\response{Table \ref{tab:models} shows the results of these fitting algorithms for different combinations of input variables and different model complexities. Large values of LOO $R^2$  indicate that the model accounts for a large portion of the variance without overfitting. Similarly, a small value of LOO $MAE$ indicates that the average difference between measured and predicted effect sizes is small. To simulate how one would assess accuracy on new data, these values are calculated after removing predictions that fall outside the range of measured effect sizes. The number of predicted effect sizes removed is also shown in Table \ref{tab:models}.  A low value of AIC indicates that the model fits well to the training data, but this may be an effect of overfitting if the model performs poorly in leave-one-out metrics. Notably, we do not claim that more complex models are not more descriptive than the models we ultimately choose; rather, our model complexity is limited by the size of our dataset.}

\response{The LOO tests also allowed us to further refine the variables to include in the final predictive model. We found that $CG$ has more predictive power than $CQ$ and therefore $CQ$ was not used in analysis (rows 1 and 2 in Table \ref{tab:models}).  We also found that using $PQ$ instead of $OG$ and $WG$  decreased predictive accuracy and so we exclude $PQ$ from the model (row 3 of Table \ref{tab:models}). Other work (e.g., \citet{weir_small_2019}) combines group work into one variable, and therefore we tried all combinations of combining group work variables. None of the combinations performed as well as including them separately (rows 8 -11 of Table \ref{tab:models}). We find that removing $OG$  improved $R^2$ and $MAE$. Our final combination of COPUS codes used as inputs is $Lec$, $CG$, $WG$, and $SQ$ (row 4 of Table \ref{tab:models}). As shown in the last 4 rows of Table \ref{tab:models}, less complex models decrease prediction accuracy substantially. }

\subsection{Out-of-Sample Predictions}
\label{subsec:bootstrap}

After determining the appropriate input variables and model complexity---a model that uses $Lec$, $CG$, $WG$, $SQ$, their squares, cubes, and quartics, and all second and third order cross terms---we performed three tests to assess how well the model can predict effect sizes for data outside the training sample. First, as discussed above and shown in Figure \ref{fig:LOO}, we used a type of data partitioning called a Leave-One-Out (LOO) test, where all data points (in this case classes) except one are used in training. \response{Feature selection is performed independently for each LOO test before bootstrapping as described in Section \ref{subsec:Modeling} of the Methods. }The model is then used to predict the effect size in the left out class to assess how well the model can predict effect sizes for classes outside of its training set. 
When we remove the \response{9} predictions that are outside the range of measured effect sizes, the mean absolute error is 0.26 and the $R^2$ value is 0.49. This means that, on average, the model makes predictions within 0.26 of the measured effect size in a given class when that class is not included in the model's training data and accounts for 49\% of the variance in effect sizes.

\begin{figure}[htbp]
    \centering
    \includegraphics[width=1\linewidth]{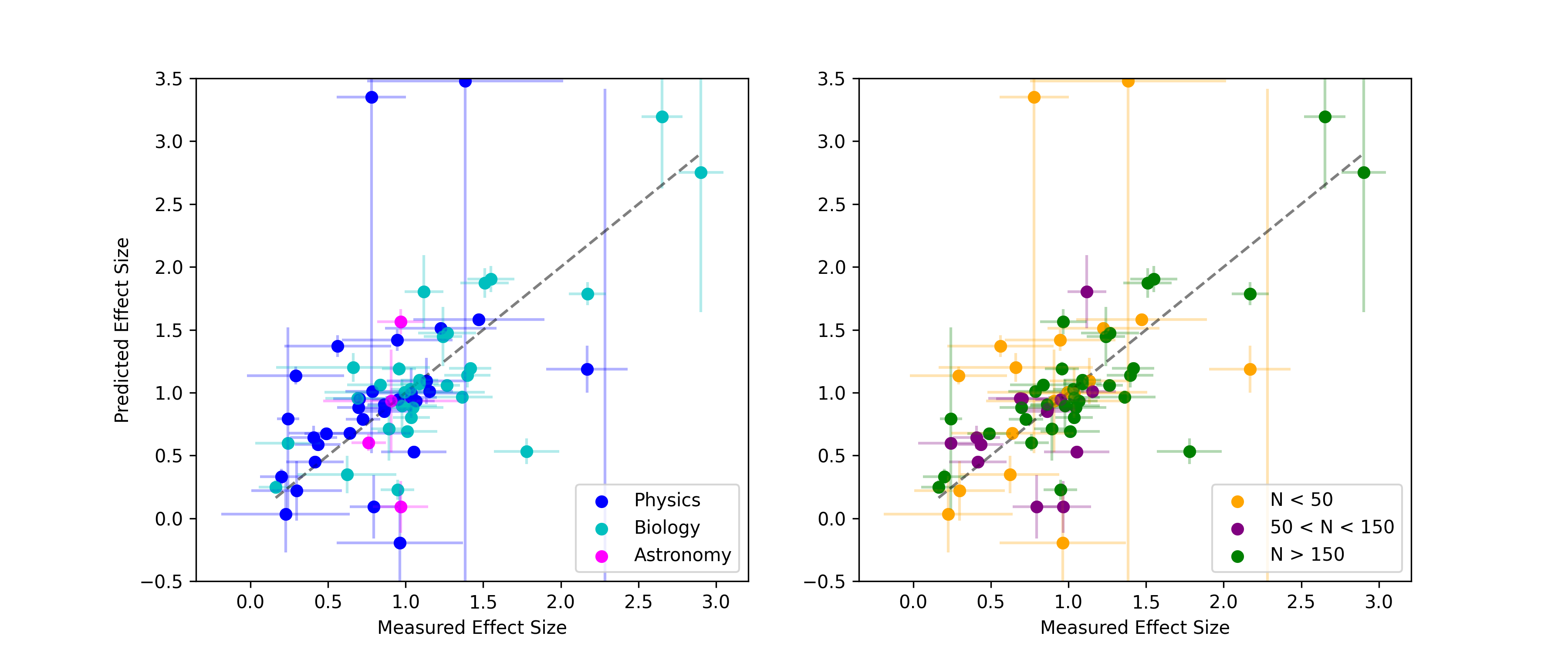}
    
    \caption{Results of Leave-One-Out tests for all classes by field (left) and class size (right).  Each class is removed from the dataset, the model is trained, and a prediction is made for the left-out class. }
    \label{fig:LOO_N_Field}
   
\end{figure}

Second, we evaluated the robustness of the model across class-level variables by distinguishing between the fields and class sizes 
during the LOO tests (Fig. \ref{fig:LOO_N_Field}). When we remove predictions outside the range of reliability, the mean absolute errors on predictions are 0.24, 0.29, and 0.26 for physics, biology, and astronomy respectively, indicating that prediction accuracy is not meaningfully impacted by discipline.  Similarly for class sizes, when unreliable predictions are removed, we find mean absolute errors of 0.33, 0.22, and 0.25 for class sizes with less than 50 students, between 50 and 150 students, and over 150 students, respectively.  Although these values are consistent for medium and large classes, we do see a higher value for class sizes less than 50. Because small class sizes have higher variance, it is unsurprising that a variance-weighted regression model would produce higher prediction errors in smaller class sizes. 

\begin{figure}
    \centering
    \includegraphics[width=0.7\linewidth]{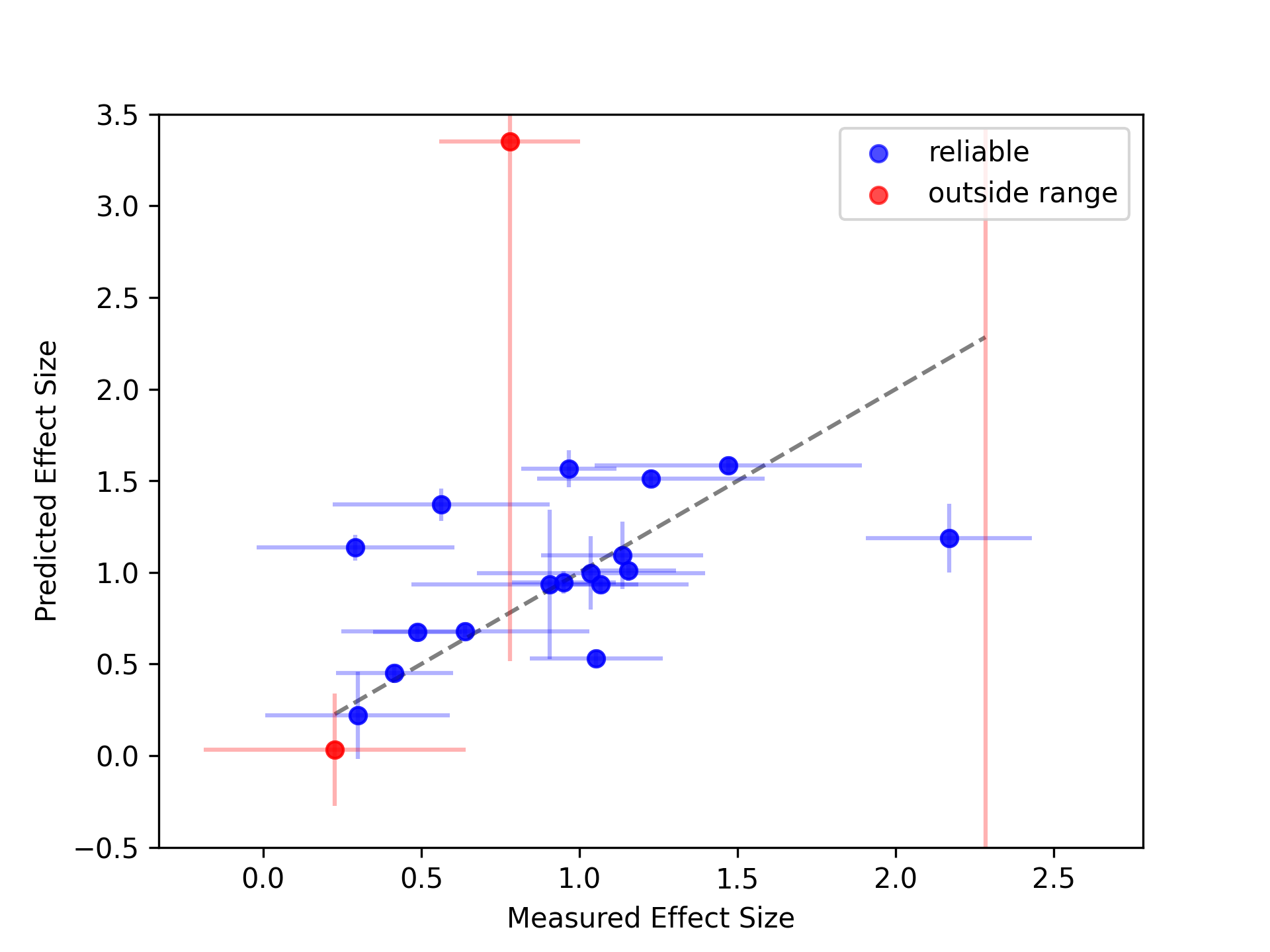}
    \caption{Results of LOO tests for classes with both unique instructors and unique institutions.}
    \label{fig:hold_out}
\end{figure}

\response{Within this dataset, there are repeated instructors and institutions. In order to make sure that overfitting is not occurring due to these repeated underlying variables, we also perform LOO tests solely on classes whose institutions and instructors are not repeated in the dataset. The results of this analysis are shown in Figure \ref{fig:hold_out}. The LOO MAE on reliable predictions for this set is 0.30, which is comparable to the analogous value in the full dataset. This indicates that there is no significant overfitting occurring due to repeated instructors or institutions.}

After ensuring that the model can reliably predict effect sizes for data outside of its training set, the model was ultimately trained on data from all 69 classes 
and used to create the predictions described in the main text (Fig. \ref{fig:heatmaps}). The number of terms included in final models ranged from 12 to 14.




    

\section{Acknowledgments}

We are grateful to the instructors who allowed data collection in their courses and the researchers who shared their data publicly. We thank the members of the first author's doctoral committee as well as Tim Steltzer, Peter Lepage, Eric Brewe, and our collaborators at Cornell University and Drexel University for their guidance and feedback at various points throughout this project. 

\section{Funding:}
O.R. discloses support for the research of this work from the U.S. National Science Foundation Graduate Research Fellowship (DGE – 2139899). N.G.H. and M.S. declare no relevant funding.

\section{Author contributions:}
O.R. conceptualized the study. O.R. and M.S. curated the data. O.R. performed the formal analysis and investigations. O.R., M.S., and N.G.H. defined the methodology.  N.G.H. and O.R. managed project administration. O.R. designed the software. N.G.H. supervised the project. O.R. conducted validation and performed visualization. O.R. wrote the original draft, which was reviewed and edited by M.S. and N.G.H.

\section{Competing interests:}
There are no competing interests to declare.

\section{Data and materials availability:}
All data and source code are available at \url{https://doi.org/10.5281/zenodo.23819570}

\end{document}